\documentclass[12pt,a4paper]{article}
\usepackage{amsmath}
\usepackage{amssymb}
\usepackage{amsthm}
\usepackage{float}
\usepackage{amsfonts}
\usepackage{graphicx}
\usepackage[]{cite}
\usepackage{float}
\usepackage{subfigure}
\usepackage{verbatim}
\usepackage[left=2cm,right=2cm,top=3cm,bottom=2.5cm]{geometry}
\usepackage[numbers]{natbib}
\usepackage[utf8]{inputenc}
\def\case#1/#2{\frac{#1}{#2}}

\def \D {\tilde{\nabla}}
\def\la {\langle}
\def\ra {\rangle}
\newcommand{\sfrac}[2]{{\textstyle{#1\over#2}}}
\def \ep {\varepsilon}
\def\tl{\tilde}
\def\rd {\displaystyle{\cdot}}
\def\ts {\textstyle}

\usepackage[usenames,dvipsnames,svgnames]{xcolor}
\usepackage[colorlinks=true,
      linkcolor=red,
      urlcolor=gray,
      citecolor=blue]{hyperref}

\def\myalign#1{%
  \def\trule{\noalign{\smallskip\hrule\medskip}}
  \def\nebc{\nearrow\bigcup}
  \def\sebc{\searrow\bigcup}
  \def\pminf{{}_{-\infty}|^{+\infty}}
  \let\Inf\infty
  \def\amp{&} 
  \vbox{\mathsurround0pt\openup1\jot
    \halign{%
      &$\displaystyle##\hfil\tabskip0pt$&\amp##\tabskip1em\crcr
      \noalign{\hrule height1pt\smallskip}#1\noalign{\smallskip\hrule height1pt}\crcr}}}

\begin{document}
\begin{center}
\textbf{Perturbations with bulk viscosity in modified chaplygin gas cosmology}
\end{center}
\hfill\\
Albert Munyeshyaka$^{1}$, Praveen Kumar Dhankar$^{2}$ and Joseph Ntahompagaze$^{3}$\\
\hfill\\ 
$^{1}$Department of Physics, Mbarara University of Science and Technology, Mbarara, Uganda\;\;\; \; \;\hfill\\
$^{2}$Symbiosis Institute of Technology, Nagpur Campus, Symbiosis International (Deemed University), Pune-440008, Maharashtra, India\;\;\; \; \;\hfill\\ 
$^{3}$ Department of Physics, College of Science and Technology, University of Rwanda, Rwanda
 \\ \\\\
Correspondence:munalph@gmail.com\;\;\;\;\;\;\;\;\;\;\;\;\;\;\;\;\;\;\;\;\;\;\;\;\;\;\;\;\;\;\;\;\;\;\;\;\;\;\;\;\;\;\;\;\;\;\;\;\;\;\;\;\;\;\;\;\;\;\;\;\;\;\;\;\;\;\;\;\;\;\;\;\;\;\;
\begin{center}
\textbf{Abstract}
\end{center}
In the present work, we investigate cosmological perturbations of viscous modified chaplygin gas model. Using $1+3$ covariant formalism, we define covariant and gauge invariant gradient variables, which after the application of scalar decomposition and  harmonic decomposition techniques together with redshift transformation method, provide the energy overdensity perturbation equations in redshift space, responsible for large scale structure formation. In order to analyse the effect of the viscous modified chaplygin gas model on matter overdensity contrast, we numerically solve the perturbation equations in both long and short wavelength limits. The numerical results show that the energy overdensity contrast decays with redshift. However, the perturbations which include amplitude effects due to the viscous modified chaplygin model do differ remarkably from those in the $\Lambda$CDM. In the absence of viscous modified chaplygin model, the results reduce to those of $\Lambda$CDM.
\\
\hfill\\
\textit{keywords:} Bulk viscosity--Chaplygin gas--Covariant perturbations-- Dark energy-- Energy density.
(This Article was accepted for publication in International Journal of Geometric Methods in Modern Physics on November 17, 2024)
\section{Introduction}\label{introduction}
It is currently a well know fact that the current expansion of the universe is in an acceleration phase. This fact was confirmed by different  observational suveys such as high redshift supernova $I_{a}$, cosmic microwave background and baryonic accoustic oscillations to name but a few \cite{riess20042004apj,aldering2002overview,perlmutter2003measuring,perlmutter2012nobel,sherwin2011evidence,das2011detection,eisenstein2005detection}. \\The standard model of comsmology dubbed $\Lambda$CDM addresses a number of problems in both the early and late universe: including the flatness problem, age problem, the horizon problem just to name a few. It is also consistent with different observations such as  the origin of cosmic microwave background radiations, the formation and distribution of large scale structure, the synthesis of light elements in the universe and its expansion \cite{weinberg2013observational,silvestri2009approaches,caldwell2009physics}. However, this model shows limitation in dealing with the cosmological constant problem, the nature and source of both dark matter and dark energy problems. As a result, this encouraged a search for  alternative  approaches to tackle these problems.\\ Among the theories seeking to address for example the dark matter and dark energy problems, chaplygin gas model and its different variants such as the Original Chaplygin Gas (OCG), Generalised Chaplygin Gas (GCG), Modified
Generalised Chaplygin Gas (MGCG), Extended Chaplygin Gas (ECG) and Generalised Cosmic Chaplygin Gas (GCCG) have been considered in the literature \cite{bento2002generalized,kamenshchik2001alternative,Bilic:2001cg,Debnath:2004cd,dev2003cosmological,akbarieh2020evolution}.\\Modified chaplygin gas has been shown to present reasonable property in describing the dark sector of the universe as single component fluid that act as both dark matter in the early and dark energy in the late universe. Different works have been considering a combination of Chaplygin gas with other theories of gravity to address different issues \cite{twagirayezu2023chaplygin,sahlu2019chaplygin,elmardi2017cosmological,karami2012polytropic}.\\ On the other hand  dissipative effects including both bulk and shear viscosity play an important role in the evolution of the universe \cite{brevik2017viscous}. Shear viscosities are ignored in this work due to the assumption that the universe is isotropic. Therefore only bulk viscosity  is considered as viscous fluid to play an important role, due to its ability to act as  a negative pressure fluid responsible for dark energy components dominating the late time universe. Furthermore, it is known that the bulk  viscosity may lead to an explanation of accelerated cosmic expansion, as discussed in different bulk viscosity models \cite{padmanabhan1987viscous,maartens1995dissipative,lima1996frw,singh2005some,sharif2017interaction}. In \cite{padmanabhan1987viscous}, the authors investigated the effect of bulk viscosity on the evolution of the universe and demonstrated that the bulk viscosity can lead to inflation-like solutions. In \cite{maartens1995dissipative}, Roy maartens made a review of dissipative cosmology and demonstrated the supports of the application of the full casual theory of bulk viscosity.\\The combination of chaplygin gas or its variants and viscosity has been considered in \cite{zhai2006viscous,saadat2013time,saadat2013frw, szydlowski2020interpretation,khadekar2019frw,singh2016bianchi}. For example the work done in \cite{zhai2006viscous} authors discussed viscous generalised chaplygin gas cosmology using dynamical analysis and showed that bulk viscosity coefficients should satisfy inequalities from the point of view of dynamics. In \cite{saadat2013time}, authours considered FRW bulk viscous cosmology in the presence of modified chaplygin gas to obtain time-dependent energy density for a flat space. In the work done in \cite{saadat2013frw}, the authors considered bulk viscous effects and chaplygin gas in the FRW cosmology in flat space and discussed the stability of this theory and presented the appropriate conditions. In \cite{dhankar2018frw}, the authors studied bulk viscous cosmology by considering modified Chaplygin  gas model in the framework of (2+1) dimensinal spacetime and discussed the stability of the model by using the speed of sound. In \cite{benaoum2014modified}, the author investigated the viscous chaplygin gas cosmological model and presented the solutions for different values of the viscous parameter.\\The study of linear cosmological perturbations with bulk viscosity in modified chaplygin gas cosmology using the $1+3$ covariant formalism was not done in the literature, to the best of our knowledge, therefore it is the main focus in this manuscript. Mainly, there are two different approaches to study cosmological perturbations, namely metric approach \cite{bardeen1980gauge,kodama1984cosmological} and $1+3$ covariant approach \cite{ellis1989covariant,abebe2012covariant,sahlu2020scalar}. The main advantage of the $1+3$ covariant approach is that the perturbations defined describe true physical degrees of freedom and no physical gauge modes present. In the recent years, the consideration of the $1+3$ covariant formalism to study cosmological perturbations have been extensively considered in both GR and different alternatives to $\Lambda$CDM model. For example the work done by \cite{abebe2023perturbations} aimed to tackle the nature of large scale structure formation through cosmological perturbations around Bianchi type-$V$ spacetime background using a $1+3$ covariant formalism. In the work done in \cite{sahlu2020scalar}, they studied perturbations in a chaplygin gas cosmology using the $1+3$ covariant formalism and the results suggest that the chaplygin gas model supports the formation of cosmic structure. In the work presented in \cite{sahlu2023confronting}, the authors considered different chaplygin gas models, confront them with supernova data and studied the perturbations of such models using the $1+3$ covariant formalism using the growth structure data $f_{\sigma 8}$.  In our recent works \cite{twagirayezu2023chaplygin,munyeshyaka2021cosmological,munyeshyaka2023multifluid,munyeshyaka20231+,munyeshyaka2024covariant}, we used the $1+3$ covariant formalism to study the perturbations in modified Gauss-Bonnet gravity and in the interacting vacuum \cite{munyeshyaka2023perturbations}.
 \\ In the present work, we  define gradient variables for both matter and modified chaplygin gas-viscous fluids to study linear cosmological perturbations and check the effect of combining both modified chaplygin and bulk viscous fluid model on the growth of matter energy density contrast with redshift. For further analysis, we  analyse the obtained perturbation equations in both long and short wavelength limits. Finally for comparison purpose, we study the growth of energy density fluctuations with redshift for both GR ($\Lambda$CDM) and our model.\\ \\
The next part of this paper is organized as follows: in Section (\ref{section2}), the presentation of the mathematical framework
 is done. Section (\ref{section3}) covers the  linear evolutions equations, and presents the scalar perturbation equations whereas Section  (\ref{section4}) presents the harmonically decomposed perturbation equations in redshift space. In Sections  (\ref{section5}) and  (\ref{section6}), we present numerical results of the perturbation equations by considering  both $GR$ limit and the system of matter-viscous modified chaplygin gas fluids in both long and short wavelength limits in both dust and radiation epochs respectively.  Section (\ref{section7})  is devoted to discussions and conclusion. The adopted spacetime signature is $(- + + +)$ and unless stated otherwise, we have used the convention $8\pi G = c = 1$, where $G$ is the gravitational constant and $c$ is the speed of light.
\section{Background Field Equations}\label{section2}
 In this section, we  present the mathematical aspects to describe the cosmic evolution. In this regard,  the action  is given as \cite{abebe2023perturbations,ntahompagaze2017f,ananda2009detailed}
\begin{eqnarray}
 S= \frac{1}{2} \int d^{4}x\sqrt{-g}\Big[R+2\mathcal{L}_m\Big],
\end{eqnarray}
 where $\mathcal{L}_{m}$ is  matter Lagrangian, Kappa ($\kappa$)  is assumed to be  $1$ and $R$ is the Ricci scalar. The  Einstein equation  is presented as
  \begin{eqnarray}
   && R_{\mu\nu}-\frac{1}{2}g^{\mu\nu}R= T^{m}_{\mu\nu},
   \label{eq2}
  \end{eqnarray}
where  $T^{m}_{\mu\nu}$ is the energy momentum tensor of the matter fluid  (photons, baryons, cold dark matter, and light neutrinos), given by
$ T_{\mu \nu} = \rho u_{\mu}u_{\nu} + p h_{\mu\nu}+ q_{\mu}u_{\nu}+ q_{\nu}u_{\mu}+\pi_{\mu \nu}$
where $\rho$, $p$, $q_{\mu}$ and $\pi_{\mu\nu}$ are the energy density, isotropic pressure, heat flux and anisotropic pressure of the fluid respectively.
The quantities $\pi_{\mu\nu}$, $q^{\mu}$, $\rho$ and $p$ are reffered  to as dynamical quantities, whereas the quantities $\sigma_{\mu\nu}$, $\varpi_{\mu\nu}$, $\theta$ and $A_{\mu}$ are reffered to as kinematical quantities. In addition to the above equations, a general model of modified chaplygin gas is given by
\begin{eqnarray}
 p=A\rho-\frac{B}{\rho^{\alpha}},
\end{eqnarray}
where $A$, $B$ and $\alpha$ are arbitrary positive constants. The MCG  interpolates from matter-dominated era to
a cosmological constant-dominated era.
On the other hand, in the study of cosmology, shear viscosities are ignored since the CMB does not indicate signiﬁcant anisotropies, and only bulk viscosities are taken into account for the viscous ﬂuid.
The bulk viscous coefficient is assumed
to have a power law dependence on the energy density and it is given by \cite{benaoum2014modified}
\begin{eqnarray}
 \xi=\xi_{0}\rho^{v},
\end{eqnarray}
where $\xi_{0}$ and $v$ are constants.
The Friedman equations which govern the evolution of the scale factor are given by
\begin{eqnarray}
 &&3H^{2}=\rho_{m}+\rho_{cv},\\
 &&3H^{2}-2\dot{H}=p_{m}+p_{cv},
\end{eqnarray}
where $cv$ stands for a combination of modified chaplygin gas and viscosity.
The consideration of  standard matter fluids (dust, radiation, etc), modified-Chaplygin gas fluid and viscosity contributions leads  us to define  the total energy density, isotropic  pressures,  as
\begin{equation}
 \rho_{t}=\rho_{m}+\rho_{cv},~ p_{t}=p_{m}+p_{cv},
 \label{eq7}
\end{equation}
where
\begin{eqnarray}
 && \rho_{cv}=\Big[\frac{B}{A+1-\sqrt{3}\xi_{0}}+\frac{C}{a^{3(\alpha+1)(A+1-\sqrt{3}\xi_{0})}}\Big]^{\frac{1}{\alpha+1}},
 \label{eq8}\\
 && p_{cv}=A\rho_{cv}-\frac{B}{\rho_{cv}^{\alpha}}-3H\xi_{0}\rho_{cv}^{\frac{1}{2}} ,\label{eq9}
 \end{eqnarray}
 where we have used $v=\frac{1}{2}$ as discussed in \cite{benaoum2014modified}. Eq. (\ref{eq8}) and Eq. (\ref{eq9}) are taken in the work presented in \cite{benaoum2014modified}.
The pressure and energy density of matter fluid are related as \begin{equation}
     p_{m}=w\rho_{m}.\label{eq10}
    \end{equation}
We assume a spatially flat Friedman-Robert-Walker(FRW)  universe,
\begin{equation}
 ds^{2}=-dt^{2}+a^{2}\left(dx^{2}+dy^{2}+dz^{2}\right),
\end{equation}
  so that the Ricci scalar is presented as
\begin{eqnarray}
   R=6(\dot{H}+2H^{2})\;,
  \label{eq12}
\end{eqnarray}
where $A$, $B$, $\alpha$ and $C$ are arbitrary constants, $H=\frac{\dot{a}}{a}=\frac{\theta}{3}$ is the Hubble parameter, $\theta$ is related to the volume expansion and $a$ is the scale factor.
The continuity equations are presented as
\begin{eqnarray}
 && \dot{\rho}_{m}=-\theta \Big(\rho_{m}+p_{m}\Big)=-\theta \Big(1+w_{m}\Big)\rho_{m}\label{eq13}\\
 && \dot{\rho}_{cv}=-3H\Big[\Big(A+1\Big)\rho_{cv}-\frac{B}{\rho^{\alpha}_{cv}}-3H\xi_{0}\rho^{\frac{1}{2}}_{cv}\Big],\label{eq14}
\end{eqnarray} where we have used the equation of state $p=w\rho$, $w$ is the equation of state parameter which is $0$ for dust matter and $w=\frac{1}{3}$ for radiation.  The Raychaudhuli equation is presented as
\begin{eqnarray}
 \dot{\theta}=-\frac{1}{3}\theta^{2}-\frac{1}{2}\Big(\rho_{t}+3p_{t}\Big)+\tilde{\bigtriangledown}^{a}\dot{u}_{a}.\label{eq15}
\end{eqnarray}
\section{The 1+3 covariant perturbation equations}\label{section3}
The $1 + 3$ covariant decomposition is a framework used in describing the linear evolution of the cosmological perturbations \cite{sahlu2020scalar,munyeshyaka20231+}.
In this approach, a fundamental observer divides space-time into hyper-surfaces and a perpendicular 4-velocity field vector where $1 + 3$ indicates the number of dimensions involved in each slice \cite{clarkson2007covariant}. That is to mean that manifold geometry of the GR is discribed in four dimensional space (ie,. time and space). It is  currently  a well known fact that the universe is not perfectly smooth, but made of large scale structures such as galaxies, clusters, voids to name but a few, believed  to be seeded from primordial fluctuations. Cosmological perturbations theory provides the mechanism to explain how these primordial fluctuations grow and form the large scale structures we see today in the universe \cite{abebe2023perturbations}. The covariant formalism describes spacetime through covariantly defined  variables with respect to the frame such as $1+3$ spacetime decomposition technique which helps in describing physics and geometry using tensor quantities and relations valid in all coordinate systems.  One of the importance of the $1 + 3$ covariant approach is to identity a set of covariant variables which describe the inhomogeneity and anisotropy of the universe \cite{ntahompagaze2018study}. In this context, we define a four-vector coordinateas function of cosmological time ($x^{\mu}=x^{\mu}(\tau)$) that  labels the comoving distance along a world-line and the corresponding velocitiy given as :
\begin{equation}
 u^{\mu}= \frac{dx^\mu}{d\tau}
\end{equation}
The projection tensor, $h_{\alpha\beta}$ into the three dimensional and orthogonal to $u^{\mu}$, satisfy the following conditions:
\begin{eqnarray}
 && h_{\alpha\beta}=g_{\alpha\beta}+u_{\alpha}u_{\beta}\Rightarrow h^{\alpha}_{\beta}h^{\beta}_{\gamma}=h^{\alpha}_{\gamma},\\
 &&h^{\alpha}_{\alpha}= 3, h_{\alpha\beta}u^{\beta}=0.
\end{eqnarray}
  The covariant derivative of the four-velocity in terms of its kinematic quantities \cite{ntahompagaze2020multifluid} is given by:
\begin{eqnarray}
 \tilde{\bigtriangledown}_{a}u_{b}= \frac{1}{3}h_{ab}\tilde{\theta} +\tilde{\sigma}_{ab}+\tilde{\omega}_{ab}-u_{a}{\dot{\tilde{u}}}_{b}
\end{eqnarray}
Where $\theta$, $\tilde{\sigma}_{ab}$, $\tilde{\omega}_{ab}$, $\dot{\tilde{u}}_{b}$, are: the volume expansion, shear tensor, vorticity tensor and four-acceleration respectively. The Hubble parameter is related to $\theta$ as $\theta=3H$. Assume the fluids in our  consideration are irrotational (ie.,$\tilde{\omega}_{ab} =0$) and shear-free (i.e $\tilde{\sigma}_{ab}=0$), the rate of expansion is given by the Raychaudhuli and conservation  equations as:
\begin{eqnarray}
 && \dot{\rho_{t}}= -\theta(\rho_{t}+p_{t}),\label{eq20}\\
 &&\tilde{\bigtriangledown}_{a}p_{t}-(\rho_{t}+p_{t})\dot{u}_{a}=0.\label{eq21}
\end{eqnarray}
Eq. (\ref{eq20})--(\ref{eq21}) are useful for constructing perturbation equations from the gradient variables of different fluids, playing key role in large structure formation.
In the following part, we assume  non-interacting  matter fluid with both modified chaplygin  gas and viscous fluids in the entire Universe where the growth of the energy overdensity fluctuations contribute to the large scale structure formation. We start by defining the covariant and Gauge-Invariant gradiant variables that describe the matter, modified chaplygin gas-viscous energy densities and expansion as per the $1+3$ covariant perturbation formalism \cite{sahlu2020scalar,abebe2012covariant,munyeshyaka2023perturbations,abebe2023perturbations}. We consider an homogenous and expanding (FRW) cosmological background to define the spatial gradient  of the Gauge invariant variables such as
\begin{eqnarray}
 && D^{m}_{a}=\frac{a \tilde{\nabla}^{a}\rho_{m}}{\rho_{m}},\label{eq22}\\
 &&Z_{a}=a\tilde{\nabla}^{a}\theta,\label{eq23}\\
 &&D_{a}^{cv}=\frac{a \tilde{\nabla}_{a}\rho_{cv}}{\rho_{cv}}\;,\label{eq24}
\end{eqnarray}
The subscipts $m$ and  $cv$  stand for matter, modified chaplygin gas-viscous fluids contributions respectively. where $D^{m}_{a}$ is the spatial gradient variable responsible for the evolution of matter overdensity, $Z_{a}$ is the gradient variable for the volume expansion and $D_{a}^{cv}$ is the gradient variable responsible for the evolution of the modified chaplygin gas-bulk viscosity energy density, respectively. Using eq. (\ref{eq7}), eq. (\ref{eq8}), eq. (\ref{eq9}) and eq. (\ref{eq10}), the $4$- velocity in eq. (\ref{eq21}) is given by
\begin{eqnarray}
 \dot{u}_{a}=-\frac{\tilde{\bigtriangledown}^{a}{p_{t}}}{\rho_{t}+p_{t}}=-\frac{1}{\rho_{t}+p_{t}}\Big[w\tilde{\bigtriangledown}^{a}\rho_{m}+\Big(A+\frac{\alpha B}{\rho^{\alpha-1}_{cv}}+\frac{3\xi_{0}H}{2\rho^{\frac{1}{2}}_{cv}}\Big)\tilde{\bigtriangledown}^{a}\rho_{cv}-\xi_{0}\rho^{\frac{1}{2}}_{cv}\tilde{\bigtriangledown}^{a}\theta \Big].\label{eq25}
\end{eqnarray}
The linear covariant identity for any scalar quantity $f$ is given by
\begin{equation}
 \Big(\tilde{\bigtriangledown}^{a}f\Big)\dot{}=\tilde{\bigtriangledown}^{a}\Big(\dot{f}\Big)-\frac{\theta}{3}\tilde{\bigtriangledown}^{a}f+\dot{f}\dot{u}_{a}.\label{eq26}
\end{equation}
The eq. (\ref{eq25}) and eq. (\ref{eq26}) are useful when constructing perturbation equations in the following ways:
The time derivative of eq. (\ref{eq22}) and by using eq. (\ref{eq10}), eq. (\ref{eq13}),  eq. (\ref{eq25}) and eq. (\ref{eq26}) then by inserting eq. (\ref{eq22})--(\ref{eq24}) yield
\begin{eqnarray}
 &&\dot{D}^{m}_{a}=\Big[\frac{\theta\Big(1+w_{m}\Big)w_{m}\rho_{m}}{\rho_{t}+p_{t}}\Big]D^{m}_{a}-\Big[(1+w_{m})\Big(1+\frac{\theta \xi_{0}\rho^{\frac{1}{2}}_{cv}}{\rho_{t}+p_{t}}\Big)\Big]Z_{a}\nonumber\\&&+\Big[\frac{\Big(A+\alpha B\rho^{1-\alpha}_{cv}+\frac{3}{2}\xi_{0}H\rho^{\frac{-1}{2}}_{cv}\Big)\rho_{cv}}{\rho_{t}+p_{t}}\Big]D^{cv}_{a}.\label{eq27}
 \end{eqnarray}
 The eq. (\ref{eq27}) is the perturbation equation for the matter overdensity. $\dot{D}^{m}_{a}$ couples with $Z_{a}$ and $D^{cv}_{a}$. The time derivative of eq. (\ref{eq23}) and by using eq. (\ref{eq8}), eq. (\ref{eq9}), eq. (\ref{eq10}), eq. (\ref{eq14}), eq. (\ref{eq15}), eq. (\ref{eq25}) and eq. (\ref{eq26}) then inserting eq. (\ref{eq22})-- (\ref{eq24}) yield
 \begin{eqnarray}
 &&\dot{Z}_{a}=\Big[-\frac{2\theta}{3}+\frac{3}{2}\xi_{0}\rho^{\frac{1}{2}}_{cv}-\Big[\frac{\theta^{2}}{3}+\frac{1}{2}\Big(1+3w_{m}\Big)\rho_{m}-\frac{1}{2}\Big(1+3A\Big)\rho_{cv}+\frac{3B}{2\rho^{\alpha}}\nonumber\\&&+\frac{9}{2} H\xi_{0}\rho^{\frac{1}{2}}\Big]\frac{\xi_{0}\rho^{\frac{1}{2}}_{cv}}{\rho_{t}+p_{t}}\Big]Z_{a}+\Big[-\frac{1}{2}\Big(1+3w_{m}\Big)\rho_{m}+\Big[\frac{\theta^{2}}{3}+\frac{1}{2}\Big(1+3w_{m}\Big)\rho_{m}-\frac{1}{2}\Big(1+3A\Big)\rho_{cv}\nonumber\\&&+\frac{3B}{2\rho^{\alpha}_{cv}}+\frac{9}{2}H\xi_{0}\rho^{\frac{1}{2}}_{cv}\Big]\frac{w_{m}\rho_{m}}{\rho_{t}+p_{t}}\Big]D^{m}_{a}+\Big[-\frac{1}{2}\Big(1+3A\big)\rho_{cv}+\frac{3\alpha B}{2\rho^{\alpha}_{cv}}+\frac{9}{2}\xi_{0}H\rho^{\frac{1}{2}}_{cv}+\Big[\frac{\theta^{2}}{3}\nonumber\\&&+\frac{1}{2}\Big(1+3w_{m}\Big)\rho_{m}-\frac{1}{2}\Big(1+3A\Big)\rho_{cv}+\frac{3B}{2\rho^{\alpha}_{cv}}+\frac{9}{2}H\xi_{0}\rho^{\frac{1}{2}}_{cv}\Big]\frac{A\rho_{cv}+\alpha B\rho^{2-\alpha}_{cv}+\frac{3}{2}\xi_{0}H\rho^{\frac{1}{2}}_{cv}}{\rho_{t}+p_{t}}\Big]D^{cv}_{a}\nonumber\\&&-\frac{w_{m}\rho_{m}}{\rho_{t}+p_{t}}\bigtriangledown^{2}D^{m}_{a}+\frac{\xi_{0}\rho^{\frac{1}{2}}_{cv}}{\rho_{t}+p_{t}}\bigtriangledown^{2}Z_{a}-\frac{A\rho_{cv}+\alpha B\rho^{2-\alpha}_{cv}+\frac{3}{2}\xi_{0}H\rho^{\frac{1}{2}}_{cv}}{\rho_{t}+p_{t}}\bigtriangledown^{2}D^{cv}_{a}\label{eq28},
 \end{eqnarray}
 eq. (\ref{eq28}) is the perturbation equation, resulting from the volume expansion. $\dot{Z}_{a}$ couples with the $D^{m}_{a}$ and $D^{cv}_{a}$. The time derivatve of eq. (\ref{eq24}) by using  eq. (\ref{eq10}), eq. (\ref{eq14}), eq. (\ref{eq25}) and eq. (\ref{eq26}) then eq. (\ref{eq22})-- (\ref{eq24}) yield
 \begin{eqnarray}
 &&\dot{D}^{cv}_{a}=\Big[-\frac{3BH}{\rho^{\alpha}_{cv}}\big(\alpha\rho_{cv}+\frac{1}{\rho_{cv}}\Big)+3H\Big(\Big(A+1\Big)\rho_{cv}-B\rho^{-\alpha}_{cv}\nonumber\\&&-3H\xi_{0}\rho^{\frac{1}{2}}_{cv}\Big)\frac{\Big(A\rho_{cv}+\alpha B\rho^{-\alpha}_{cv}+\frac{3}{2}\xi_{0}H\rho^{\frac{1}{2}}_{cv}\Big)}{\rho_{t}+p_{t}}\Big]D^{cv}_{a}+\Big[3H\Big(\Big(A+1\Big)\rho_{cv}-B\rho^{-\alpha}_{cv}\nonumber\\&&-3H\xi_{0}\rho^{\frac{1}{2}}_{cv}\Big)\frac{w_{m}\rho_{m}}{\rho_{t}+p_{t}}\Big]D^{m}_{a}+\Big[-\Big(A+1\Big)+B\rho^{-(\alpha+1)}_{cv}+2\xi_{0}\theta\rho^{-\frac{1}{2}}_{cv}-3H\Big(\Big(A+1\Big)\rho_{cv}\nonumber\\&&-B\rho^{-\alpha}_{cv}-3H\xi_{0}\rho^{\frac{1}{2}}_{cv}\Big)\frac{\xi_{0}\rho^{\frac{1}{2}}_{cv}}{\rho_{t}+p_{t}}\Big]Z_{a}. \label{eq29}
\end{eqnarray}
The eq. (\ref{eq29}) is the perturbation  equation of the Chaplygin gas-viscous fluids.  $\dot{D}^{cv}_{a}$ couples with $Z_{a}$ and $D^{m}_{a}$. We notice that the system of evolution equations (eq. (\ref{eq27})--(\ref{eq29})) form a closed system of first-order partial differential equations responsible for the evolution of perturbations. Our main interest lies in the large structure formation and it is generally believed that the large scale structure formation  follows a spherical clustering \cite{abebe2023perturbations,munyeshyaka2023perturbations,carloni2008evolution}. Therefore only scalar part of perturbation equations plays a key role. The spherically symmetric components of the vector gradient variables can be extracted from the defined vector gradient variables using local decomposition and scalar decomposition techniques and presented as
\begin{eqnarray}
 &&\Delta^{m}=\tilde{\nabla}^{a}\Big({\frac{a \tilde{\nabla}^{a}\rho_{m}}{\rho_{m}}\Big)},\label{eq30}\\
 &&Z=\tilde{\nabla}^{a}\Big(a\tilde{\nabla}^{a}\theta\Big),\label{eq31}\\
 &&\Delta^{cv}=\tilde{\nabla}^{a}\Big(\frac{a \tilde{\nabla}_{a}\rho_{cv}}{\rho_{cv}}\Big)\;.\label{eq32}
 \end{eqnarray}
  The scalar parts (eq. (\ref{eq30})--(\ref{eq32})) evolve as
  \begin{eqnarray}
 &&\dot{\Delta}_{m}=\Big[\frac{\theta\Big(1+w_{m}\Big)w_{m}\rho_{m}}{\rho_{t}+p_{t}}\Big]\Delta_{m}-\Big[(1+w_{m})\Big(1+\frac{\theta \xi_{0}\rho^{\frac{1}{2}}_{cv}}{\rho_{t}+p_{t}}\Big)\Big]Z\nonumber\\&&+\Big[\frac{\Big(A+\alpha B\rho^{1-\alpha}_{cv}+\frac{3}{2}\xi_{0}H\rho^{\frac{-1}{2}}_{cv}\Big)\rho_{cv}}{\rho_{t}+p_{t}}\Big]\Delta_{cv},\label{eq33}\\
 &&\dot{Z}=\Big[-\frac{2\theta}{3}+\frac{3}{2}\xi_{0}\rho^{\frac{1}{2}}_{cv}-\Big[\frac{\theta^{2}}{3}+\frac{1}{2}\Big(1+3w_{m}\Big)\rho_{m}-\frac{1}{2}\Big(1+3A\Big)\rho_{cv}+\frac{3B}{2\rho^{\alpha}}\nonumber\\&&+\frac{9}{2} H\xi_{0}\rho^{\frac{1}{2}}\Big]\frac{\xi_{0}\rho^{\frac{1}{2}}_{cv}}{\rho_{t}+p_{t}}\Big]Z+\Big[-\frac{1}{2}\Big(1+3w_{m}\Big)\rho_{m}+\Big[\frac{\theta^{2}}{3}+\frac{1}{2}\Big(1+3w_{m}\Big)\rho_{m}-\frac{1}{2}\Big(1+3A\Big)\rho_{cv}\nonumber\\&&+\frac{3B}{2\rho^{\alpha}_{cv}}+\frac{9}{2}H\xi_{0}\rho^{\frac{1}{2}}_{cv}\Big]\frac{w_{m}\rho_{m}}{\rho_{t}+p_{t}}\Big]\Delta_{m}+\Big[-\frac{1}{2}\Big(1+3A\big)\rho_{cv}+\frac{3\alpha B}{2\rho^{\alpha}_{cv}}\nonumber\\&&+\frac{9}{2}\xi_{0}H\rho^{\frac{1}{2}}_{cv}+\Big[\frac{\theta^{2}}{3}+\frac{1}{2}\Big(1+3w_{m}\Big)\rho_{m}-\frac{1}{2}\Big(1+3A\Big)\rho_{cv}\nonumber\\&&+\frac{3B}{2\rho^{\alpha}_{cv}}+\frac{9}{2}H\xi_{0}\rho^{\frac{1}{2}}_{cv}\Big]\frac{A\rho_{cv}+\alpha B\rho^{2-\alpha}_{cv}+\frac{3}{2}\xi_{0}H\rho^{\frac{1}{2}}_{cv}}{\rho_{t}+p_{t}}\Big]\Delta_{cv}-\frac{w_{m}\rho_{m}}{\rho_{t}+p_{t}}\bigtriangledown^{2}\Delta_{m}+\frac{\xi_{0}\rho^{\frac{1}{2}}_{cv}}{\rho_{t}+p_{t}}\bigtriangledown^{2}Z\nonumber\\
 &&-\frac{A\rho_{cv}+\alpha B\rho^{2-\alpha}_{cv}+\frac{3}{2}\xi_{0}H\rho^{\frac{1}{2}}_{cv}}{\rho_{t}+p_{t}}\bigtriangledown^{2}\Delta_{cv}\\\label{eq34}
 &&\dot{\Delta}_{cv}=\Big[-\frac{3BH}{\rho^{\alpha}_{cv}}\big(\alpha\rho_{cv}+\frac{1}{\rho_{cv}}\Big)+3H\Big(\Big(A+1\Big)\rho_{cv}-B\rho^{-\alpha}_{cv}\nonumber\\&&-3H\xi_{0}\rho^{\frac{1}{2}}_{cv}\Big)\frac{\Big(A\rho_{cv}+\alpha B\rho^{-\alpha}_{cv}+\frac{3}{2}\xi_{0}H\rho^{\frac{1}{2}}_{cv}\Big)}{\rho_{t}+p_{t}}\Big]\Delta_{cv}+\Big[3H\Big(\Big(A+1\Big)\rho_{cv}-B\rho^{-\alpha}_{cv}-3H\xi_{0}\rho^{\frac{1}{2}}_{cv}\Big)\frac{w_{m}\rho_{m}}{\rho_{t}+p_{t}}\Big]\Delta_{m}\nonumber\\&&+\Big[-\Big(A+1\Big)+B\rho^{-(\alpha+1)}_{cv}+2\xi_{0}\theta\rho^{-\frac{1}{2}}_{cv}-3H\Big(\Big(A+1\Big)\rho_{cv}-B\rho^{-\alpha}_{cv}-3H\xi_{0}\rho^{\frac{1}{2}}_{cv}\Big)\frac{\xi_{0}\rho^{\frac{1}{2}}_{cv}}{\rho_{t}+p_{t}}\Big]Z\label{eq35}
\end{eqnarray}
The equations eq. (\ref{eq33})--(\ref{eq35}) are the evolution of perturbations responsible for large scale structure formation. Rewritting eq. (\ref{eq33})--(\ref{eq35}) as a system of ordinary differential equations, we need to use the harmonic decomposition technique.
\section{Harmonic decomposition}\label{section4}
This technique is commonly used to convert partial differential equations into ordinary differential equations as pointed out in \cite{sahlu2020scalar,ntahompagaze2018study,abebe2015breaking,murorunkwere20211+}.  We write the covariantly defined Laplace-Beltrami operator as
\begin{eqnarray}
 \tilde{\bigtriangledown}^{2}Y=-\frac{k^{2}}{a^{2}}Y,
\end{eqnarray}
where $Y$ are the eigenfunctions of the covariant operator, $k$ is the wave-number related to the cosmological scale factor as $k=\frac{2\pi a}{\lambda}$, with $\lambda$ is the wavelength of perturbation. Thus the evolution of perturbation equations eq. (\ref{eq33})--(\ref{eq35}) in the $k^{th}$ mode can be rewritten in harmonic space as
\begin{eqnarray}
 &&\dot{\Delta}^{k}_{m}=\Big[\frac{\theta\Big(1+w_{m}\Big)w_{m}\rho_{m}}{\rho_{t}+p_{t}}\Big]\Delta^{k}_{m}-\Big[(1+w_{m})\Big(1+\frac{\theta \xi_{0}\rho^{\frac{1}{2}}_{cv}}{\rho_{t}+p_{t}}\Big)\Big]Z^{k}\nonumber\\&&+\Big[\frac{\Big(A+\alpha B\rho^{1-\alpha}_{cv}+\frac{3}{2}\xi_{0}H\rho^{\frac{-1}{2}}_{cv}\Big)\rho_{cv}}{\rho_{t}+p_{t}}\Big]\Delta^{k}_{cv},\label{eq37}\\
 &&\dot{Z}^{k}=\Big[-\frac{2\theta}{3}+\frac{3}{2}\xi_{0}\rho^{\frac{1}{2}}_{cv}-\Big[\frac{\theta^{2}}{3}+\frac{1}{2}\Big(1+3w_{m}\Big)\rho_{m}-\frac{1}{2}\Big(1+3A\Big)\rho_{cv}+\frac{3B}{2\rho^{\alpha}}+\frac{9}{2} H\xi_{0}\rho^{\frac{1}{2}}\Big]\frac{\xi_{0}\rho^{\frac{1}{2}}_{cv}}{\rho_{t}+p_{t}}\nonumber\\&&-\frac{\xi_{0}\rho^{\frac{1}{2}}_{cv}}{\rho_{t}+p_{t}}\frac{k^{2}}{a^{2}}\Big]Z^{k}+\Big[-\frac{1}{2}\Big(1+3w_{m}\Big)\rho_{m}+\Big[\frac{\theta^{2}}{3}+\frac{1}{2}\Big(1+3w_{m}\Big)\rho_{m}-\frac{1}{2}\Big(1+3A\Big)\rho_{cv}+\frac{3B}{2\rho^{\alpha}_{cv}}\nonumber\\&&+\frac{9}{2}H\xi_{0}\rho^{\frac{1}{2}}_{cv}\Big]\frac{w_{m}\rho_{m}}{\rho_{t}+p_{t}}+\frac{w_{m}\rho_{m}}{\rho_{t}+p_{t}}\frac{k^{2}}{a^{2}}\Big]\Delta^{k}_{m}+\Big[-\frac{1}{2}\Big(1+3A\big)\rho_{cv}+\frac{3\alpha B}{2\rho^{\alpha}_{cv}}+\frac{9}{2}\xi_{0}H\rho^{\frac{1}{2}}_{cv}\nonumber\\&&+\Big[\frac{\theta^{2}}{3}+\frac{1}{2}\Big(1+3w_{m}\Big)\rho_{m}-\frac{1}{2}\Big(1+3A\Big)\rho_{cv}+\frac{3B}{2\rho^{\alpha}_{cv}}+\frac{9}{2}H\xi_{0}\rho^{\frac{1}{2}}_{cv}\Big]\frac{A\rho_{cv}+\alpha B\rho^{2-\alpha}_{cv}+\frac{3}{2}\xi_{0}H\rho^{\frac{1}{2}}_{cv}}{\rho_{t}+p_{t}}\nonumber\\&&+\frac{A\rho_{cv}+\alpha B\rho^{2-\alpha}_{cv}+\frac{3}{2}\xi_{0}H\rho^{\frac{1}{2}}_{cv}}{\rho_{t}+p_{t}}\frac{k^{2}}{a^{2}}\Big]\Delta^{k}_{cv},\label{eq38}\\
 &&\dot{\Delta}^{k}_{cv}=\Big[-\frac{3BH}{\rho^{\alpha}_{cv}}\big(\alpha\rho_{cv}+\frac{1}{\rho_{cv}}\Big)+3H\Big(\Big(A+1\Big)\rho_{cv}-B\rho^{-\alpha}_{cv}\nonumber\\&&-3H\xi_{0}\rho^{\frac{1}{2}}_{cv}\Big)\frac{\Big(A\rho_{cv}+\alpha B\rho^{-\alpha}_{cv}+\frac{3}{2}\xi_{0}H\rho^{\frac{1}{2}}_{cv}\Big)}{\rho_{t}+p_{t}}\Big]\Delta^{k}_{cv}+\Big[3H\Big(\Big(A+1\Big)\rho_{cv}-B\rho^{-\alpha}_{cv}-3H\xi_{0}\rho^{\frac{1}{2}}_{cv}\Big)\frac{w_{m}\rho_{m}}{\rho_{t}+p_{t}}\Big]\Delta^{k}_{m}\nonumber\\&&+\Big[-\Big(A+1\Big)+B\rho^{-(\alpha+1)}_{cv}+2\xi_{0}\theta\rho^{-\frac{1}{2}}_{cv}-3H\Big(\Big(A+1\Big)\rho_{cv}-B\rho^{-\alpha}_{cv}-3H\xi_{0}\rho^{\frac{1}{2}}_{cv}\Big)\frac{\xi_{0}\rho^{\frac{1}{2}}_{cv}}{\rho_{t}+p_{t}}\Big]Z^{k}, \label{eq39}
\end{eqnarray}
The eq. (\ref{eq37})--(\ref{eq39}) are ordinary differential equations which are much easier to handle. By using the redshift transformation scheme defined below, \cite{twagirayezu2023chaplygin,sahlu2020scalar,munyeshyaka2023perturbations,hough2021confronting}
  \begin{eqnarray*}
&& a=\frac{1}{1+z},\\
&& \dot{f}=-(1+z)Hf',\\
&& \ddot{f}=(1+z)^{2}H\Big(\frac{dH}{dz}\frac{df}{dz}+H\frac{d^{2}f}{dz^{2}}\Big)+(1+z)H^{2}\frac{df}{dz},
\end{eqnarray*}
we can express the perturbation equations in the redshift space as
\begin{eqnarray}
 &&-\Big(1+z\Big)H\Delta'^{k}_{m}=\Big[\frac{\theta\Big(1+w_{m}\Big)w_{m}\rho_{m}}{\rho_{t}+p_{t}}\Big]\Delta^{k}_{m}-\Big[(1+w_{m})\Big(1+\frac{\theta \xi_{0}\rho^{\frac{1}{2}}_{cv}}{\rho_{t}+p_{t}}\Big)\Big]Z^{k}\nonumber\\&&+\Big[\frac{\Big(A+\alpha B\rho^{1-\alpha}_{cv}+\frac{3}{2}\xi_{0}H\rho^{\frac{-1}{2}}_{cv}\Big)\rho_{cv}}{\rho_{t}+p_{t}}\Big]\Delta^{k}_{cv},\label{eq40}\\
 &&-\Big(1+z\Big)HZ'^{k}=\Big[-\frac{2\theta}{3}+\frac{3}{2}\xi_{0}\rho^{\frac{1}{2}}_{cv}-\Big[\frac{\theta^{2}}{3}+\frac{1}{2}\Big(1+3w_{m}\Big)\rho_{m}-\frac{1}{2}\Big(1+3A\Big)\rho_{cv}+\frac{3B}{2\rho^{\alpha}}\nonumber\\&&+\frac{9}{2} H\xi_{0}\rho^{\frac{1}{2}}\Big]\frac{\xi_{0}\rho^{\frac{1}{2}}_{cv}}{\rho_{t}+p_{t}}-\frac{\xi_{0}\rho^{\frac{1}{2}}_{cv}}{\rho_{t}+p_{t}}\frac{k^{2}}{a^{2}}\Big]Z^{k}+\Big[-\frac{1}{2}\Big(1+3w_{m}\Big)\rho_{m}+\Big[\frac{\theta^{2}}{3}+\frac{1}{2}\Big(1+3w_{m}\Big)\rho_{m}-\frac{1}{2}\Big(1+3A\Big)\rho_{cv}\nonumber\\&&+\frac{3B}{2\rho^{\alpha}_{cv}}+\frac{9}{2}H\xi_{0}\rho^{\frac{1}{2}}_{cv}\Big]\frac{w_{m}\rho_{m}}{\rho_{t}+p_{t}}+\frac{w_{m}\rho_{m}}{\rho_{t}+p_{t}}\frac{k^{2}}{a^{2}}\Big]\Delta^{k}_{m}+\Big[-\frac{1}{2}\Big(1+3A\big)\rho_{cv}+\frac{3\alpha B}{2\rho^{\alpha}_{cv}}+\frac{9}{2}\xi_{0}H\rho^{\frac{1}{2}}_{cv}\nonumber\\&&+\Big[\frac{\theta^{2}}{3}+\frac{1}{2}\Big(1+3w_{m}\Big)\rho_{m}-\frac{1}{2}\Big(1+3A\Big)\rho_{cv}+\frac{3B}{2\rho^{\alpha}_{cv}}+\frac{9}{2}H\xi_{0}\rho^{\frac{1}{2}}_{cv}\Big]\frac{A\rho_{cv}+\alpha B\rho^{2-\alpha}_{cv}+\frac{3}{2}\xi_{0}H\rho^{\frac{1}{2}}_{cv}}{\rho_{t}+p_{t}}\nonumber\\&&+\frac{A\rho_{cv}+\alpha B\rho^{2-\alpha}_{cv}+\frac{3}{2}\xi_{0}H\rho^{\frac{1}{2}}_{cv}}{\rho_{t}+p_{t}}\frac{k^{2}}{a^{2}}\Big]\Delta^{k}_{cv},\label{eq41}\\
 &&-\Big(1+z\Big)H\Delta'^{k}_{cv}=\Big[-\frac{3BH}{\rho^{\alpha}_{cv}}\big(\alpha\rho_{cv}+\frac{1}{\rho_{cv}}\Big)+3H\Big(\Big(A+1\Big)\rho_{cv}-B\rho^{-\alpha}_{cv}\nonumber\\&&-3H\xi_{0}\rho^{\frac{1}{2}}_{cv}\Big)\frac{\Big(A\rho_{cv}+\alpha B\rho^{-\alpha}_{cv}+\frac{3}{2}\xi_{0}H\rho^{\frac{1}{2}}_{cv}\Big)}{\rho_{t}+p_{t}}\Big]\Delta^{k}_{cv}+\Big[3H\Big(\Big(A+1\Big)\rho_{cv}-B\rho^{-\alpha}_{cv}\nonumber\\&&-3H\xi_{0}\rho^{\frac{1}{2}}_{cv}\Big)\frac{w_{m}\rho_{m}}{\rho_{t}+p_{t}}\Big]\Delta^{k}_{m}+\Big[-\Big(A+1\Big)+B\rho^{-(\alpha+1)}_{cv}+2\xi_{0}\theta\rho^{-\frac{1}{2}}_{cv}\nonumber\\
 &&-3H\Big(\Big(A+1\Big)\rho_{cv}-B\rho^{-\alpha}_{cv}-3H\xi_{0}\rho^{\frac{1}{2}}_{cv}\Big)\frac{\xi_{0}\rho^{\frac{1}{2}}_{cv}}{\rho_{t}+p_{t}}\Big]Z^{k}.\label{eq42}
\end{eqnarray}
The perturbation equations eq. (\ref{eq40})--(\ref{eq42}) are closed and can now be solved numerically by considering long and short wave limits.  The main aim of this work is to check if there is any effect of generalized chaplygin gas-viscous fluid mixture on the growth rate of matter density perturbations. We set the initial conditions at some redshift $z_{in}$ (for example $z_{in}=4$) to solve the system of equations for $\Delta^{k}_{m}(z)$ and make a comparison  with the $\Lambda$CDM limit.
For $\Lambda$CDM limit, that is to say $A=0$, $B=0$, $\xi_{0}=0$, eq. (\ref{eq40})--(\ref{eq42}) are reduced to
\begin{eqnarray}
 &&-\Big(1+z\Big)H\Delta'^{k}_{m}=\Big(\theta w_{m}\Big)\Delta^{k}_{m}-\Big[(1+w_{m})\Big]Z^{k},\label{eq43}\\
 &&-\Big(1+z\Big)HZ'^{k}=\Big(-\frac{2\theta}{3}\Big)Z^{k}+\Big[-\frac{\Big(1+3w_{m}\Big)\rho_{m}}{2\Big(1+w_{m}\Big)}+\Big(\frac{\theta^{2}}{3}+\frac{k^{2}}{a^{2}}\Big)\frac{w_{m}}{1+w_{m}}\Big]\Delta^{k}_{m}, \label{eq44}\\
 &&\Delta'^{k}_{cv}=0\label{eq45}.
\end{eqnarray}
Define the matter overdensity contrast as \cite{sahlu2020scalar,sami2021covariant}
\begin{equation}
 \delta(z)=\frac{\Delta^{k}(z)}{\Delta^{k}(z_{in})},
\end{equation}
with $z_{in}=4$ in both $\Lambda$CDM and our current model. We have also used the following relations: $\rho_{m}=\rho_{m0}(1+z)^{3}$ for dust dominated universe ($w=0$) and $\rho_{m}=\rho_{m0}(1+z)^{4}$ for radiation dominated universe ($w=\frac{1}{3}$). Next sections analyse the perturbations for a dust or radiation dominated universe in the long and short wavelength limits.
\section{Perturbations in long wavelength limints}\label{section5}
Further analysis of perturbation equations in long wavelength limit is made by considering that the wave-number $k$ is much smaller compared to other terms in the perturbations \cite{sahlu2020scalar,abebe2012covariant}. In this limit, we study the evolution of energy density perturbations in both dust and radiation dominated universe.
\subsection{Dust dominated Universe}
\begin{figure}
\includegraphics[width=100mm, height=60mm]{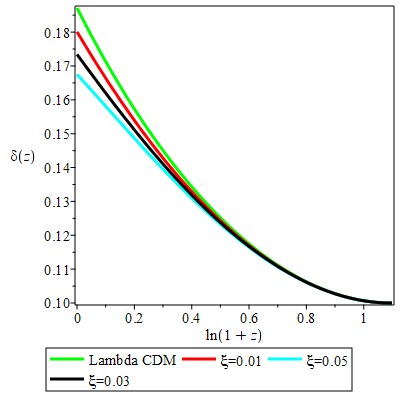}
\caption{Plot of energy density contrast vs redshift of equations eq. (\ref{eq47})--eq. (\ref{eq49}) using $A=\frac{1}{3}$, $B=1$ and $C=1$ in the dust dominated universe ($w=0$), long Wavelength limit ($k=0.000001$), $\Delta^{m}(z_{in})=10^{-5}$, $Z_{in}=10^{-5}$ and $\Delta^{cv}(z_{in})=10^{-5}$ were used as initial conditions.}
\label{fig1}
\end{figure}

\begin{figure}
\includegraphics[width=100mm, height=60mm]{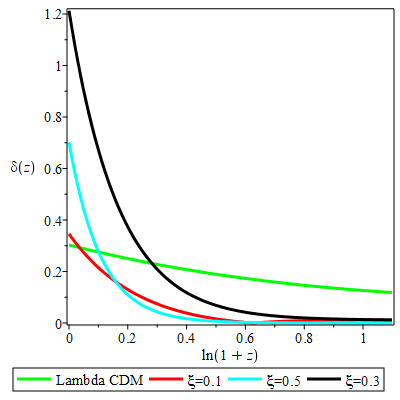}
\caption{Plot of energy density contrast vs redshift of equations eq. (\ref{eq47})--eq. (\ref{eq49})  $A=\frac{1}{3}$, $B=1$ and $C=1$ in the dust dominated universe ($w=0$), long Wavelength limit ($k=0.000001$), $\Delta^{m}(z_{in})=10^{-5}$, $Z_{in}=10^{-5}$ and $\Delta^{cv}(z_{in})=10^{-5}$ were used as initial conditions.}\label{fig2}
\end{figure}

\begin{figure}
\includegraphics[width=100mm, height=60mm]{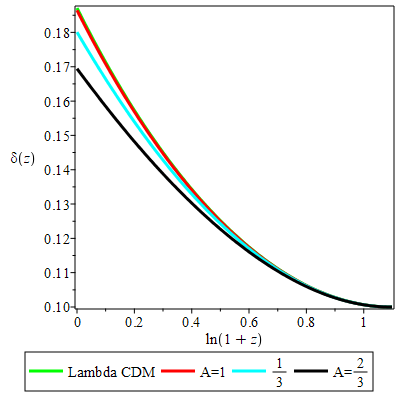}
\caption{Plot of energy density contrast vs redshift of equations eq. (\ref{eq47})--eq. (\ref{eq49})  $\xi=0.1$ , $B=1$ and $C=1$ in the dust dominated universe ($w=0$), long Wavelength limit ($k=0.000001$), $\Delta^{m}(z_{in})=10^{-5}$, $Z_{in}=10^{-5}$ and $\Delta^{cv}(z_{in})=10^{-5}$ were used as initial conditions.}\label{fig3}
\end{figure}

The perturbation equations are analysed for the dust dominated universe, where the equation of state parameter is set to $w=0$. The perturbation equations eq. (\ref{eq40})--(\ref{eq42}) can then be rewritten as
\begin{eqnarray}
 &&-\Big(1+z\Big)H\Delta'^{k}_{d}=-\Big[\Big(1+\frac{\theta \xi_{0}\rho^{\frac{1}{2}}_{cv}}{\rho_{t}+p_{t}}\Big)\Big]Z^{k}+\Big[\frac{\Big(A+\alpha B\rho^{1-\alpha}_{cv}+\frac{3}{2}\xi_{0}H\rho^{\frac{-1}{2}}_{cv}\Big)\rho_{cv}}{\rho_{t}+p_{t}}\Big]\Delta^{k}_{cv},\label{eq47}\\
 &&-\Big(1+z\Big)HZ'^{k}=\Big[-\frac{2\theta}{3}+\frac{3}{2}\xi_{0}\rho^{\frac{1}{2}}_{cv}-\Big[\frac{\theta^{2}}{3}+\frac{1}{2}\rho_{d}-\frac{1}{2}\Big(1+3A\Big)\rho_{cv}+\frac{3B}{2\rho^{\alpha}}+\frac{9}{2} H\xi_{0}\rho^{\frac{1}{2}}\Big]\frac{\xi_{0}\rho^{\frac{1}{2}}_{cv}}{\rho_{t}+p_{t}}\nonumber\\&&-\frac{\xi_{0}\rho^{\frac{1}{2}}_{cv}}{\rho_{t}+p_{t}}\frac{k^{2}}{a^{2}}\Big]Z^{k}-\Big(\frac{1}{2}\rho_{d}\Big)\Delta^{k}_{m}+\Big[-\frac{1}{2}\Big(1+3A\big)\rho_{cv}+\frac{3\alpha B}{2\rho^{\alpha}_{cv}}+\frac{9}{2}\xi_{0}H\rho^{\frac{1}{2}}_{cv}\nonumber\\&&+\Big[\frac{\theta^{2}}{3}+\frac{1}{2}\rho_{d}-\frac{1}{2}\Big(1+3A\Big)\rho_{cv}+\frac{3B}{2\rho^{\alpha}_{cv}}+\frac{9}{2}H\xi_{0}\rho^{\frac{1}{2}}_{cv}\Big]\frac{A\rho_{cv}+\alpha B\rho^{2-\alpha}_{cv}+\frac{3}{2}\xi_{0}H\rho^{\frac{1}{2}}_{cv}}{\rho_{t}+p_{t}}\nonumber\\&&+\frac{A\rho_{cv}+\alpha B\rho^{2-\alpha}_{cv}+\frac{3}{2}\xi_{0}H\rho^{\frac{1}{2}}_{cv}}{\rho_{t}+p_{t}}\frac{k^{2}}{a^{2}}\Big]\Delta^{k}_{cv},\label{eq48}\\
 &&-\Big(1+z\Big)H\Delta'^{k}_{cv}=\Big[-\frac{3BH}{\rho^{\alpha}_{cv}}\big(\alpha\rho_{cv}+\frac{1}{\rho_{cv}}\Big)+3H\Big(\Big(A+1\Big)\rho_{cv}-B\rho^{-\alpha}_{cv}\nonumber\\&&-3H\xi_{0}\rho^{\frac{1}{2}}_{cv}\Big)\frac{\Big(A\rho_{cv}+\alpha B\rho^{-\alpha}_{cv}+\frac{3}{2}\xi_{0}H\rho^{\frac{1}{2}}_{cv}\Big)}{\rho_{t}+p_{t}}\Big]\Delta^{k}_{cv}+\Big[-\Big(A+1\Big)+B\rho^{-(\alpha+1)}_{cv}+2\xi_{0}\theta\rho^{-\frac{1}{2}}_{cv}\nonumber\\
 &&-3H\Big(\Big(A+1\Big)\rho_{cv}-B\rho^{-\alpha}_{cv}-3H\xi_{0}\rho^{\frac{1}{2}}_{cv}\Big)\frac{\xi_{0}\rho^{\frac{1}{2}}_{cv}}{\rho_{t}+p_{t}}\Big]Z^{k}.\label{eq49}
\end{eqnarray}
Solving eq. (\ref{eq47})--eq. (\ref{eq49}) numerically, the results are presented in fig. (\ref{fig1}), fig. (\ref{fig2}) and fig. (\ref{fig3}). This is done by changing the parameters $\xi$ or $A$ and by fixing $B$ and $C$.
\\By looking at both fig. (\ref{fig1}), fig. (\ref{fig2}) and fig. (\ref{fig3}), the matter energy density contrast decays with redshift but with amplitudes higher than that of $\Lambda$CDM limit, for the considered values of viscosity parameter $\xi_{0}$ and $A$.
\subsection{Radiation dominated Universe}
\begin{figure}
\includegraphics[width=100mm, height=60mm]{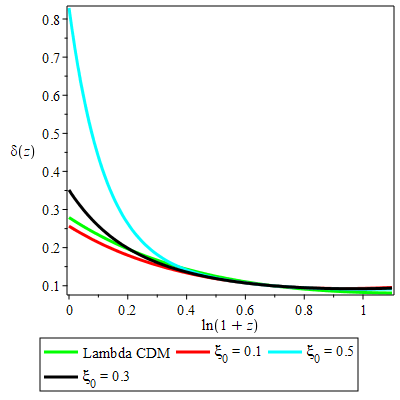}
\caption{Plot of energy density contrast vs redshift of equations eq. (\ref{eq50})--eq. (\ref{eq52}) using $A=\frac{1}{3}$, $B=1$ and $C=1$ in the radiation dominated universe ($w=\frac{1}{3}$), long Wavelength limit ($k=0$), $\Delta^{m}(z_{in})=10^{-5}$, $Z_{in}=10^{-5}$ and $\Delta^{cv}(z_{in})=10^{-5}$ were used as initial conditions.}\label{fig4}
\end{figure}

\begin{figure}
\includegraphics[width=100mm, height=60mm]{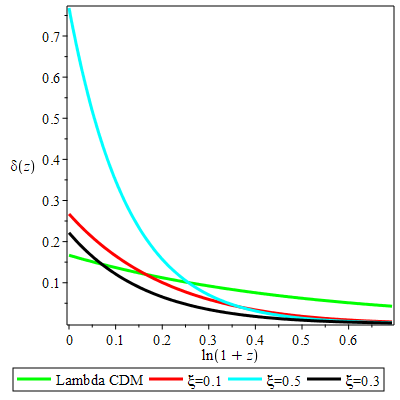}
\caption{Plot of energy density contrast vs redshift of equations eq. (\ref{eq50})--eq. (\ref{eq52})  using $A=\frac{1}{3}$, $B=1$ and $C=1$ in the radiation dominated universe ($w=\frac{1}{3}$), long Wavelength limit ($k=0.000001$), $\Delta^{m}(z_{in})=10^{-5}$, $Z_{in}=10^{-5}$ and $\Delta^{cv}(z_{in})=10^{-5}$ were used as initial conditions.}\label{fig5}
\end{figure}

\begin{figure}
\includegraphics[width=100mm, height=60mm]{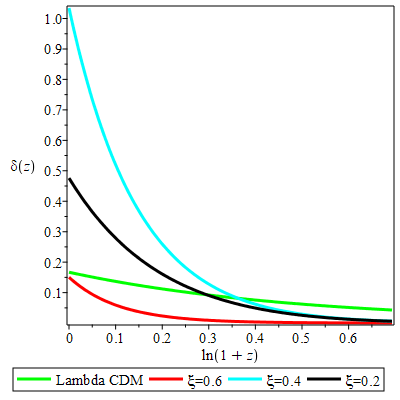}
\caption{Plot of energy density contrast vs redshift of equations eq. (\ref{eq50})--eq. (\ref{eq52}), using $A=\frac{1}{3}$, $B=1$ and $C=1$ in the radiation dominated universe ($w=\frac{1}{3}$), long Wavelength limit ($k=0.0001$), $\Delta^{m}(z_{in})=10^{-5}$, $Z_{in}=10^{-5}$ and $\Delta^{cv}(z_{in})=10^{-5}$ were used as initial conditions.}\label{fig6}
\end{figure}

\begin{figure}
\includegraphics[width=100mm, height=60mm]{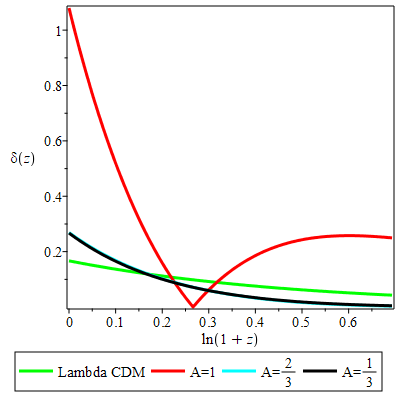}
\caption{Plot of energy density contrast vs redshift of equations eq. (\ref{eq50})--eq. (\ref{eq52}), using $\xi=0.1$, $B=1$ and $C=1$ in the radiation dominated universe ($w=\frac{1}{3}$), long Wavelength limit ($k=0.0001$), $\Delta^{m}(z_{in})=10^{-5}$, $Z_{in}=10^{-5}$ and $\Delta^{cv}(z_{in})=10^{-5}$ were used as initial conditions.}\label{fig7}
\end{figure}

By assuming that the universe is dominated by radiation, it means the equation of state parameter is set to $w=\frac{1}{3}$, eq. (\ref{eq40})--(\ref{eq42}) are then rewritten by
\begin{eqnarray}
 &&-\Big(1+z\Big)H\Delta'^{k}_{r}=\frac{4\theta\rho_{r}}{9\Big(\rho_{t}+p_{t}\Big)}\Delta^{k}_{r}-\Big[\frac{4}{3}\Big(1+\frac{\theta \xi_{0}\rho^{\frac{1}{2}}_{cv}}{\rho_{t}+p_{t}}\Big)\Big]Z^{k}\nonumber\\&&+\Big[\frac{\Big(A+\alpha B\rho^{1-\alpha}_{cv}+\frac{3}{2}\xi_{0}H\rho^{\frac{-1}{2}}_{cv}\Big)\rho_{cv}}{\rho_{t}+p_{t}}\Big]\Delta^{k}_{cv},\label{eq50}\\
 &&-\Big(1+z\Big)HZ'^{k}=\Big[-\frac{2\theta}{3}+\frac{3}{2}\xi_{0}\rho^{\frac{1}{2}}_{cv}-\Big[\frac{\theta^{2}}{3}+\rho_{r}-\frac{1}{2}\Big(1+3A\Big)\rho_{cv}+\frac{3B}{2\rho^{\alpha}}+\frac{9}{2} H\xi_{0}\rho^{\frac{1}{2}}\Big]\frac{\xi_{0}\rho^{\frac{1}{2}}_{cv}}{\rho_{t}+p_{t}}\nonumber\\&&-\frac{\xi_{0}\rho^{\frac{1}{2}}_{cv}}{\rho_{t}+p_{t}}\frac{k^{2}}{a^{2}}\Big]Z^{k}+\Big[-\rho_{r}+\Big[\frac{\theta^{2}}{3}+\rho_{r}-\frac{1}{2}\Big(1+3A\Big)\rho_{cv}+\frac{3B}{2\rho^{\alpha}_{cv}}\nonumber\\&&+\frac{9}{2}H\xi_{0}\rho^{\frac{1}{2}}_{cv}\Big]\frac{\rho_{r}}{3\Big(\rho_{t}+p_{t}\Big)}+\frac{\rho_{r}}{3\Big(\rho_{t}+p_{t}\Big)}\frac{k^{2}}{a^{2}}\Big]\Delta^{k}_{r}+\Big[-\frac{1}{2}\Big(1+3A\big)\rho_{cv}+\frac{3\alpha B}{2\rho^{\alpha}_{cv}}+\frac{9}{2}\xi_{0}H\rho^{\frac{1}{2}}_{cv}\nonumber\\&&+\Big[\frac{\theta^{2}}{3}+\rho_{r}-\frac{1}{2}\Big(1+3A\Big)\rho_{cv}+\frac{3B}{2\rho^{\alpha}_{cv}}+\frac{9}{2}H\xi_{0}\rho^{\frac{1}{2}}_{cv}\Big]\frac{A\rho_{cv}+\alpha B\rho^{2-\alpha}_{cv}+\frac{3}{2}\xi_{0}H\rho^{\frac{1}{2}}_{cv}}{\rho_{t}+p_{t}}\nonumber\\&&+\frac{A\rho_{cv}+\alpha B\rho^{2-\alpha}_{cv}+\frac{3}{2}\xi_{0}H\rho^{\frac{1}{2}}_{cv}}{\rho_{t}+p_{t}}\frac{k^{2}}{a^{2}}\Big]\Delta^{k}_{cv},\label{eq51}\\
 &&-\Big(1+z\Big)H\Delta'^{k}_{cv}=\Big[-\frac{3BH}{\rho^{\alpha}_{cv}}\big(\alpha\rho_{cv}+\frac{1}{\rho_{cv}}\Big)+3H\Big(\Big(A+1\Big)\rho_{cv}-B\rho^{-\alpha}_{cv}\nonumber\\&&-3H\xi_{0}\rho^{\frac{1}{2}}_{cv}\Big)\frac{\Big(A\rho_{cv}+\alpha B\rho^{-\alpha}_{cv}+\frac{3}{2}\xi_{0}H\rho^{\frac{1}{2}}_{cv}\Big)}{\rho_{t}+p_{t}}\Big]\Delta^{k}_{cv}+\Big[3H\Big(\Big(A+1\Big)\rho_{cv}-B\rho^{-\alpha}_{cv}\nonumber\\&&-3H\xi_{0}\rho^{\frac{1}{2}}_{cv}\Big)\frac{\rho_{r}}{3\Big(\rho_{t}+p_{t}\Big)}\Big]\Delta^{k}_{r}+\Big[-\Big(A+1\Big)+B\rho^{-(\alpha+1)}_{cv}+2\xi_{0}\theta\rho^{-\frac{1}{2}}_{cv}\nonumber\\
 &&-3H\Big(\Big(A+1\Big)\rho_{cv}-B\rho^{-\alpha}_{cv}-3H\xi_{0}\rho^{\frac{1}{2}}_{cv}\Big)\frac{\xi_{0}\rho^{\frac{1}{2}}_{cv}}{\rho_{t}+p_{t}}\Big]Z^{k}. \label{eq52}
\end{eqnarray}
Solving eq. (\ref{eq50})-eq. (\ref{eq52}) numerically, numerical results of energy density \cite{dent2011f} are presented in fig. (\ref{fig4}), fig. (\ref{fig5}), fig. (\ref{fig6}) and fig. (\ref{fig7}).
Dynamical system analysis \cite{basilakos2019dynamical,khyllep2207cosmology} can be used for constraining different considered parameters
\section{Perturbations in short wavelength limits}\label{section6}
\begin{figure}
\includegraphics[width=100mm, height=60mm]{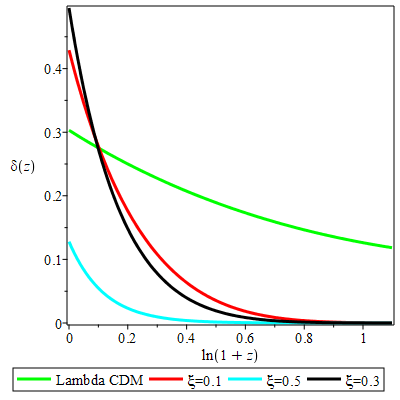}
\caption{Plot of energy density contrast vs redshift of equations eq. (\ref{eq47})--eq. (\ref{eq49}), using $A=\frac{1}{3}$, $B=1$ and $C=1$  in the dust ($w=0$) dominated universe, short wavelength ($k=10$) limit, $\Delta^{m}(z_{in})=10^{-5}$, $Z_{in}=10^{-5}$ and $\Delta^{cv}(z_{in})=10^{-5}$ have been used as initial conditions.}\label{fig8}
\end{figure}

\begin{figure}
\includegraphics[width=100mm, height=60mm]{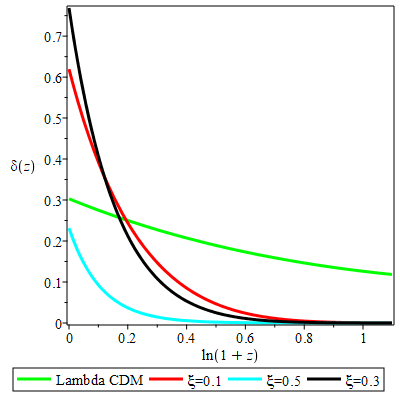}
\caption{Plot of energy density contrast vs redshift of equations eq. (\ref{eq47})--eq. (\ref{eq49}), using $A=\frac{1}{3}$, $B=1$ and $C=1$  in the dust ($w=0$) dominated universe, short wavelength ($k=100$) limit, $\Delta^{m}(z_{in})=10^{-5}$, $Z_{in}=10^{-5}$ and $\Delta^{cv}(z_{in})=10^{-5}$ have been used as initial conditions.}\label{fig9}
\end{figure}

\begin{figure}
\includegraphics[width=100mm, height=60mm]{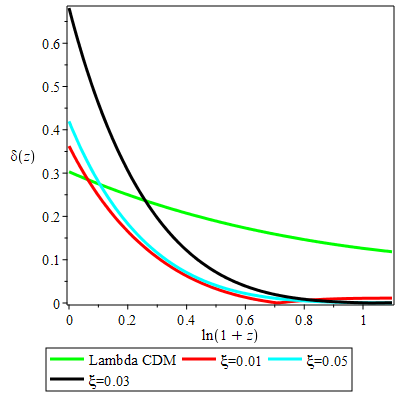}
\caption{Plot of energy density contrast vs redshift of equations eq. (\ref{eq47})--eq. (\ref{eq49}), using $A=\frac{1}{3}$, $B=1$ and $C=1$  in the dust ($w=0$) dominated universe, short wavelength ($k=10$) limit, $\Delta^{m}(z_{in})=10^{-5}$, $Z_{in}=10^{-5}$ and $\Delta^{cv}(z_{in})=10^{-5}$ have been used as initial conditions.}\label{fig10}
\end{figure}

\begin{figure}
\includegraphics[width=100mm, height=60mm]{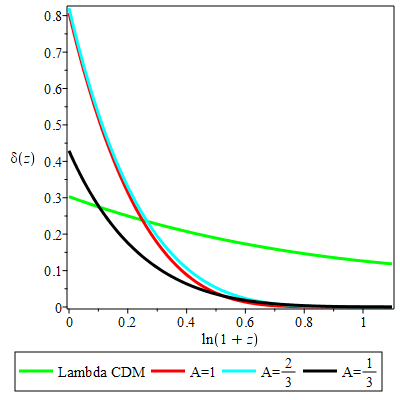}
\caption{Plot of energy density contrast vs redshift of equations eq. (\ref{eq47})--eq. (\ref{eq49}), using $\xi=0.1$, $B=1$ and $C=1$  in the dust ($w=0$) dominated universe, short wavelength ($k=100$) limit, $\Delta^{m}(z_{in})=10^{-5}$, $Z_{in}=10^{-5}$ and $\Delta^{cv}(z_{in})=10^{-5}$ have been used as initial conditions.}\label{fig11}
\end{figure}

\begin{figure}
\includegraphics[width=100mm, height=60mm]{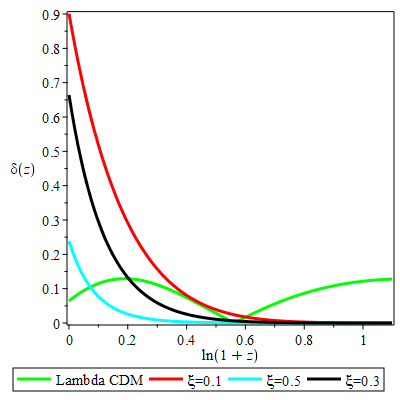}
\caption{Plot of energy density contrast vs redshift of equations eq. (\ref{eq50})-eq. (\ref{eq52}), using $A=\frac{1}{3}$, $B=1$ and $C=1$  in the radiation ($w=\frac{1}{3}$) dominated universe, short wavelength ($k=100$) limit,$\Delta^{m}(z_{in})=10^{-5}$, $Z_{in}=10^{-5}$ and $\Delta^{cv}(z_{in})=10^{-5}$ have been used as initial conditions.}\label{fig12}
\end{figure}

\begin{figure}
\includegraphics[width=100mm, height=60mm]{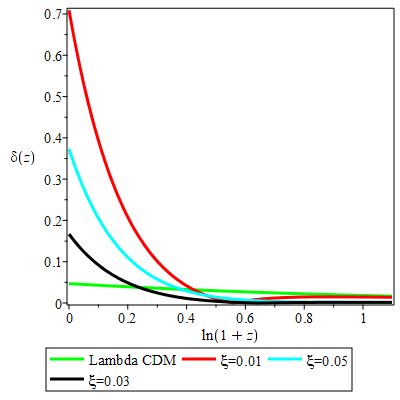}
\caption{Plot of energy density contrast vs redshift of equations eq. (\ref{eq50})-eq. (\ref{eq52}), using $A=\frac{1}{3}$, $B=1$ and $C=1$  in the radiation ($w=\frac{1}{3}$) dominated universe, short wavelength ($k=10$) limit, $\Delta^{m}(z_{in})=10^{-5}$, $Z_{in}=10^{-5}$ and $\Delta^{cv}(z_{in})=10^{-5}$ have been used as initial conditions}.\label{fig13}
\end{figure}

\begin{figure}
\includegraphics[width=100mm, height=60mm]{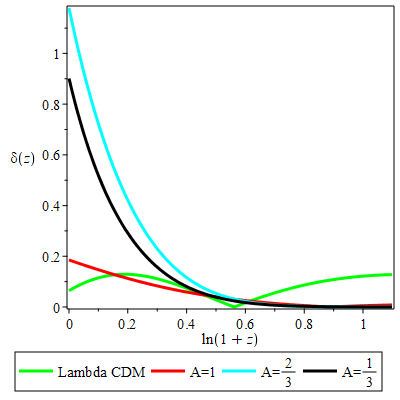}
\caption{Plot of energy density contrast vs redshift of equations eq. (\ref{eq50})-eq. (\ref{eq52}), using $\xi=0.1$, $B=1$ and $C=1$  in the radiation ($w=\frac{1}{3}$) dominated universe, short wavelength ($k=100$) limit, $\Delta^{m}(z_{in})=10^{-5}$, $Z_{in}=10^{-5}$ and $\Delta^{cv}(z_{in})=10^{-5}$ have been used as initial conditions}.\label{fig14}
\end{figure}

The short wavelength mode assumes that the value of the wave-number $k$ is very large compared to other terms in the perturbations \cite{sahlu2020scalar, munyeshyaka2023perturbations}. Applying this approximation, we analyse the perturbations in both dust and radiation dominated universe as we did in the long wavelength limit and present numerical results.
By considering that the $k$ is very large compared to other terms in eq. (\ref{eq47})--eq. (\ref{eq49}) in the dust dominated case and solving numerically, the numerical results are presented in fig. (\ref{fig8}), fig. (\ref{fig9}), fig. (\ref{fig10}) and fig. (\ref{fig11}) for different values of viscosity constant $\xi_{0}$ and parameter $A$.
 By considering different values of $k$, the perturbations in short wavelength are scale dependance, a result in agreement with the work conducted in\cite{saridakis2021we,saridakis2023observational}.
 Applying the short wavelength approximation to the eq. (\ref{eq50})-eq. (\ref{eq52}) in the radiation dominated case and solving numerically, the results are presented in fig. (\ref{fig12}), fig. (\ref{fig13}) and fig. (\ref{fig14}). 

\section{Discussions and conclusion}\label{section7}
The current treatment consider a mixture of matter, modified chaplygin gas and bulk viscous fluids model in flat homogeneous and isotropic universe to study cosmological perturbations. Using some of the basic equations governing this model presented in \cite{benaoum2014modified} together with the defined covariant gauge invariant gradient variables, we presented the perturbation equations in redshift space responsible for large scale structure formation.  In order to further analyse the effect of modified chaplygin gas-bulk viscous fluid on the matter energy overdensity, we solved numerically the perturbations equations.  We have analysed the growth of density contrast for both long and short wavelength limits. In the long wavelength limit, we  apply the assumption that the wave number $k$ is much smaller compared to other terms in the perturbations. We further consider the cases where the universe is filled with  dust fluid, where the equation state parameter is given by $w=0$ to get the equations Eq. (\ref{eq47})--(\ref{eq49}). In the second case, we assume that the universe is filled with radiation mixture, where $w=\frac{1}{3}$ to get perturbation equations Eq. (\ref{eq50})--(\ref{eq52}). Using different model parameters and initial conditions such as  $B=1$, $C=1$, $\Delta^{m}(z_{in})=10^{-5}$, $Z_{in}=10^{-5}$ and $\Delta^{cv}(z_{in})=10^{-5}$ for some $z_{in}=4$, and by changing the viscosity parameter $\xi_{0}$ or $A$. We get numerical results of perturbation equations as presented in figs. (\ref{fig1})--(\ref{fig3}) for dust dominated universe and in figs. (\ref{fig4})--(\ref{fig7}) for radiation dominated universe. From the plots, the density contrast ($\delta(z)$) decays with redshift.  In the case of short wavelength limit, we assume the wave number is much bigger than other terms in the perturbations so that the numerical results are presented in figs. (\ref{fig8})--(\ref{fig11}) for dust dominated universe whereas  fig. (\ref{fig12}), fig. (\ref{fig13}) and fig. (\ref{fig14}) represent the numerical results for radiation dominated universe. By looking at all the plots in both long and short wavelength limits, we depict that the energy density contrast ($\delta(z)$) decays with redshift for both $\Lambda$CDM limit and our model, but the amplitudes of perturbations are higher than those  of $\Lambda$CDM. The decay of density contrast suggests the possibility for large scale structure formation in the universe which may help in cosmic acceleration scenario.  Moreover, our results suggest that for both long wavelength (it means smaller values of the wave-number $k$) and short wavelength limits, as we change the value of viscosity constant or the parameter $A$, the amplitude of perturbations (spread in amplitudes) get larger in a dust dominated universe compared to those in the radiation dominated universe. By investigating the evolution of $\delta(z)$ of the viscous  modified chaplygin gas model, where $v=\frac{1}{2}$ for different viscous parameter $\xi_{0}$, one can depict that the $\delta(z)$ tends to $\Lambda$CDM as $z$ increases. As a follow up task, it is worthwhile considering the $1+3$ covariant cosmological perturbations for a general model ($v\neq \frac{1}{2}$) and/or by considering  other variants of chaplygin gas model. This will be done elsewhere. 
\section*{Acknowledgements}
AM acknowledges financial support from  the Swedish International Development Agency (SIDA) through International Science Program (ISP) to East African Astronomical Research Network (EAARN) (grant number  AFRO:05). AM also acknowledges the hospitality of the Department of Physics of the University of Rwanda, where this work was conceptualized and completed. PKD would like to thank the COST Action CA18108 "Quantum gravity phenomenology in the multi-messenger approach" for providing financial support to visit the University of Rijeka, Croatia and also PKD acknowledges IUCAA, Pune, India for giving him a visiting associateship position. JN acknowledges the financial support from SIDA through ISP to the University of Rwanda (UR) through Rwanda Astrophysics Space and Climate Science Research Group (RASCRG), Grant number RWA:$01$.
\appendix
\section{Useful Linearised Differential Identities}
For all scalars $f$, vectors $V_a$ and tensors that vanish in the background,
$S_{ab}=S_{\la ab\ra}$, the following linearised identities hold:
\begin{eqnarray}
\left(\D_{\la a}\D_{b\ra}f\right)^{.}&=&\D_{\la a}\D_{b\ra}\dot{f}-\sfrac{2}{3}\Theta\D_{\la a}\D_{b\ra}f+\dot{f}\D_{\la a}A_{b\ra}\label{a0}\;,\\
\ep^{abc}\D_b \D_cf &=& 0 \label{a1}\;, \\
\ep_{cda}\D^{c}\D_{\la b}\D^{d\ra}f&=&\ep_{cda}\D^{c}\D_{( b}\D^{d)}f=\ep_{cda}\D^{c}\D_{ b}\D^{d}f=0\label{a2}\;,\\
\D^2\left(\D_af\right) &=&\D_a\left(\D^2f\right) 
+\sfrac{1}{3}\tl{R}\D_a f \label{a4}\;,\\
\left(\D_af\right)^{\rd} &=& \D_a\dot{f}-\sfrac{1}{3}\Theta\D_af+\dot{f}A_a 
\label{a14}\;,\\
\left(\D_aS_{b\cdots}\right)^{\rd} &=& \D_a\dot{S}_{b\cdots}
-\sfrac{1}{3}\Theta\D_aS_{b\cdots}
\label{a15}\;,\\
\left(\D^2 f\right)^{\rd} &=& \D^2\dot{f}-\sfrac{2}{3}\Theta\D^2 f 
+\dot{f}\D^a A_a \label{a21}\;,\\
\D_{[a}\D_{b]}V_c &=& 
-\sfrac{1}{6}\tl{R}V_{[a}h_{b]c} \label{a16}\;,\\
\D_{[a}\D_{b]}S^{cd} &=& -\sfrac{1}{3}\tl{R}S_{[a}{}^{(c}h_{b]}{}^{d)} \label{a17}\;,\\
\D^a\left(\ep_{abc}\D^bV^c\right) &=& 0 \label{a20}\;,\\
\label{divcurl}\D_b\left(\ep^{cd\la a}\D_c S^{b\ra}_d\right) &=& {\ts{1\over2}}\ep^{abc}\D_b \left(\D_d S^d_c\right)\;,\\
\text{curlcurl} V_{a}&=&\D_{a}\left(\D^{b}V_{b}\right)-\D^{2}V_{a}+\sfrac{1}{3}\tl{R}V_{a}\label{curlcurla}\;,
\label{a21}
\end{eqnarray}
\clearpage

\bibliographystyle{iopart-num}
\providecommand{\newblock}{}

 \noindent
{\color{blue} \rule{\linewidth}{1mm} }
  \end{document}